\documentclass[pra,twocolumn,10pt,superscriptaddress,noshowpacs,floatfix]{revtex4}
\usepackage[pdftex,plainpages=false,colorlinks=true,citecolor=blue,linkcolor=blue,urlcolor=blue,filecolor=green,bookmarksopen=true]{hyperref}
\usepackage[utf8]{inputenc}
\usepackage{amsmath}
\usepackage{tikz}
\usepackage[normalem]{ulem}
\usepackage[caption = false]{subfig}
\usepackage{graphicx,epstopdf}
\usepackage{blindtext}
\usepackage{lipsum}
\usepackage{amsfonts}
\usepackage{bbm}
\usepackage{amssymb}
\usepackage{enumerate}
\usepackage{color}
\usepackage{latexsym}
\usepackage{times,txfonts}

\newcommand{\Dcal}{\mathcal{D}}

\newcommand{\Acal}{\mathcal{A}}
\newcommand{\Lcal}{\mathcal{L}}
\newcommand{\Ecal}{\mathcal{E}}

\newcommand{\Fcal}{\mathcal{F}}

\newcommand{\1}{\mathbbm{1}}
\newcommand{\Lmath}{\mathbbm{L}}
\newcommand{\Cmath}{\mathbbm{C}}

\newcommand{\dket}[1]{| #1 \rangle\rangle}
\newcommand{\dbra}[1]{\langle\langle #1 |}
\newcommand{\ket}[1]{| #1 \rangle}

\newcommand{\dinterpro}[2]{\langle \langle #1 | #2 \rangle \rangle}

\newcommand{\bra}[1]{\langle #1 |}

\newcommand{\trs}[1]{ \text{Tr}\{ #1 \}}

\DeclareFontFamily{U}{mathc}{}
\DeclareFontShape{U}{mathc}{m}{it}%
{<->s*[1.03] mathc10}{}
\DeclareMathAlphabet{\mathscr}{U}{mathc}{m}{it}

\begin{document}

\title{Adiabatic quantum dynamics under decoherence in a controllable trapped-ion setup}

\author{Chang-Kang Hu}
\affiliation{CAS Key Laboratory of Quantum Information, University of Science and Technology of China, Hefei 230026, People’s Republic of China}
\affiliation{CAS Center For Excellence in Quantum Information and Quantum Physics, University of Science and Technology of China, Hefei 230026, People’s Republic of China}

\author{Alan C. Santos}
\email{ac\_santos@id.uff.br}
\affiliation{Instituto de F\'{i}sica, Universidade Federal Fluminense, Av. Gal. Milton Tavares de Souza s/n, Gragoat\'{a}, 24210-346 Niter\'{o}i, Rio de Janeiro, Brazil}

\author{Jin-Ming Cui}
\email{jmcui@ustc.edu.cn}
\affiliation{CAS Key Laboratory of Quantum Information, University of Science and Technology of China, Hefei 230026, People’s Republic of China}
\affiliation{CAS Center For Excellence in Quantum Information and Quantum Physics, University of Science and Technology of China, Hefei 230026, People’s Republic of China}

\author{\\Yun-Feng Huang}
\email{hyf@ustc.edu.cn}
\affiliation{CAS Key Laboratory of Quantum Information, University of Science and Technology of China, Hefei 230026, People’s Republic of China}
\affiliation{CAS Center For Excellence in Quantum Information and Quantum Physics, University of Science and Technology of China, Hefei 230026, People’s Republic of China}

\author{Marcelo S. Sarandy}
\email{msarandy@id.uff.br}
\affiliation{Instituto de F\'{i}sica, Universidade Federal Fluminense, Av. Gal. Milton Tavares de Souza s/n, Gragoat\'{a}, 24210-346 Niter\'{o}i, Rio de Janeiro, Brazil}

\author{Chuan-Feng Li}
\email{cfli@ustc.edu.cn}
\affiliation{CAS Key Laboratory of Quantum Information, University of Science and Technology of China, Hefei 230026, People’s Republic of China}
\affiliation{CAS Center For Excellence in Quantum Information and Quantum Physics, University of Science and Technology of China, Hefei 230026, People’s Republic of China}

\author{Guang-Can Guo}
\affiliation{CAS Key Laboratory of Quantum Information, University of Science and Technology of China, Hefei 230026, People’s Republic of China}
\affiliation{CAS Center For Excellence in Quantum Information and Quantum Physics, University of Science and Technology of China, Hefei 230026, People’s Republic of China}

\begin{abstract}
Suppressing undesired nonunitary effects is a major challenge in quantum computation and quantum control. In this work, by considering the adiabatic dynamics in presence of a surrounding environment, we theoretically and 
experimentally analyze the robustness of adiabaticity in open quantum systems. More specifically, by considering a decohering scenario, we exploit the validity conditions of the adiabatic approximation 
as well as its sensitiveness to the resonance situation, which typically harm adiabaticity in closed systems. As an illustration, we implement an oscillating Landau-Zener Hamiltonian, which shows that decoherence may drive the
resonant system with high fidelities to the adiabatic behavior of open systems. 
Moreover we also implement the adiabatic  quantum algorithm for the Deutsch problem, where a distinction is established 
between the open system adiabatic density operator and the target pure state expected in the computation process. 
Preferred time windows for obtaining the desired outcomes are then analyzed. 
We experimentally realize these systems through a single trapped Ytterbium ion $^{171} $Yb$^+$, where the ion hyperfine energy levels 
are used as degrees of freedom of a two-level system, with both driven field and decohering strength efficiently controllable.
\end{abstract}

\maketitle

\section{Introduction}

The impossibility of decoupling an open quantum system from its surrounding environment has motivated the development of methods to minimize the
harmful nonunitary effects on a quantum evolution~\cite{Petruccione:Book}. In some specific cases (see, e.g., Ref.~\cite{Childs:01}), such effects can be smoothened for slowly driven Hamiltonians, which are governed by the adiabatic theorem~\cite{Born:28,Messiah:Book,Kato:50}. Adiabatic evolutions have been extensively
used in a number of applications, such as geometric phases \cite{Berry:84,Wilczek:84}, quantum computation~\cite{Farhi:01,Bacon:09,Barends:16,Johnson:11,Hen:15},  quantum
thermodynamics~\cite{Alicki:79,Kosloff:84,Quan:07,Geva:92,Henrich:07}, quantum game theory~\cite{dePonte:18}, among others~\cite{Vitanov:99,Amniat:12,Chu:08}.
 For closed systems, adiabaticity is associated with a decoupled dynamics of eigenspaces corresponding to distinct energy eigenvalues in the
 spectrum of a quantum Hamiltonian. In particular, assuming no level crossings in the energy spectrum along the evolution, if a quantum state is prepared in an instantaneous
 non-degenerate eigenstate of a sufficiently slowly varying Hamiltonian $H(t)$ at an initial time then it will evolve to the corresponding instantaneous eigenstate
 at later times. Remarkably, this picture has been challenged for quantum systems driven by highly oscillating fields under resonant conditions~\cite{Marzlin:04},
which may imply the failure of the expected adiabatic behavior. In particular, this issue has been experimentally investigated in a nuclear magnetic
 resonance setup~\cite{Suter:08}, where it has been shown that many previously introduced quantitative adiabatic conditions~\cite{Tong:07,Jianda-Wu:08,Ambainis:04} may
 fail at resonance. As an alternative to deal with this problem, a validation mechanism for adiabaticity
 has been recently introduced and experimentally realized via ion traps~\cite{Hu-18-b}. In this approach,
 adiabatic conditions may be consistently verified by adopting non-inertial reference frames.

Concerning open quantum systems, a generalization of the adiabatic theorem can be directly achieved for nonunitary evolution under
time-local master equations governed by a Lindbladian super-operator $\Lcal_{t}$. In this case, as originally derived in Ref.~\cite{Sarandy:05-1}, 
open system adiabaticity can be defined by replacing the concept
of decoupled evolution of eigenspaces of a Hamiltonian $H(t)$ for the decoupled evolution of Jordan blocks~\cite{Horn:Book} of a Lindbladian $\Lcal_{t}$. 
The open system adiabatic behavior reduces to the closed case as decoherence vanishes and implies, in general, in a finite optimal time for Jordan 
block decoupling, since there
is a competition between the adiabatic time scale (requiring slow evolution) and the relaxation time scale (requiring fast evolution)~\cite{Sarandy:05-2}.
This is rather different from the traditional closed system case, where adiabaticity is favored by long evolution times, being exact at the infinite time limit. In a related approach, Refs.~\cite{Venuti:16,Venuti:17} consider adiabatic paths that exactly hold in the infinite time limit for open quantum systems 
initially prepared in the steady state of $\Lcal_{t}$ at $t=0$, which is associated with the Lindblad Jordan block with zero eigenvalue.

In this work, we introduce a realization of the open system adiabatic theorem considering quantum states initially prepared in general 
one-dimensional Jordan blocks of $\Lcal_{t}$, 
investigating the adiabatic behavior in a decohering environment
both at resonance and off-resonance.
The experimental setup is based on a hyperfine quantum bit (qubit) built upon a single trapped
Ytterbium ion $^{171} $Yb$^+$, where the ion hyperfine energy levels are used as degrees of freedom of a two-level system
(see, e.g., Refs.~\cite{Hu:18,Hu-18-b}).
As illustrations of our approach, we implement an oscillating Landau-Zener Hamiltonian, specifically introduced to investigate the effects of resonance,
and the adiabatic quantum algorithm for the Deutsch problem, where fidelities can be interpreted in terms of computation outcomes. As expected in a nonunitary evolution, we will emphasize a distinction between the open system (mixed) adiabatic density operator and 
the target pure state expected as the result of the computation process. As we will show, 
since an open system typically evolves to a mixed state, the adiabatic density operator will usually provide (at most) an approximation of the target 
desired state. 
This approximation will occur at specific finite time windows, which must be taken into account to ensure the success of the algorithm. Moreover, for both cases, we provide a class of density operators for which the open system adiabaticity occurs at the infinite time limit,
circumventing the usual competition between the adiabatic and relaxation time scales. As a by-product, this is shown to simplify the general quantitative
condition obtained Ref.~\cite{Sarandy:05-1} for a class of initial quantum states.

\section{Adiabatic approximation for open quantum systems}

Let us consider a discrete $D$-dimensional open quantum system described by a density operator ${\rho}(t)$. The nonunitary evolution of the
system is assumed to be driven by a time-local master equation given by
\begin{equation}
\dot{\rho}(t) = \Lcal_{t}[\rho(t)],
\label{eq0}
\end{equation}
where $\Lcal_{t}[\bullet]$ is the Lindbladian operator and the subscript $t$ denotes that $\Lcal_{t}[\bullet]$ may be time-dependent.
As in Ref.~\cite{Sarandy:05-1}, the open system adiabatic dynamics is conveniently derived by adopting the superoperator formalism~\cite{Alicki:Book1}.
In this formalism, Eq.~(\ref{eq0}) can be rewritten as
\begin{equation}
\dket{\dot{\rho}(t)} = \Lmath(t) \dket{\rho(t)},
\end{equation}
where $\dket{\rho(t)}$ is the $D^2$-dimensional coherence vector and $\Lmath(t)$ is a $(D^2 \times D^2)$-dimensional matrix. The components $\varrho_{n}(t)$ of the vector $\dket{\rho(t)}$ are obtained from $\varrho_{n}(t) = \trs{\rho(t)\sigma_{n}^{\dagger}}$ and the elements $\Lmath_{mn}(t)$ of $\Lmath(t)$ are written as $\Lmath_{mn}(t) = (1/D)\trs{\sigma_{m}^{\dagger}\Lcal_{t}[\sigma_{n}]}$, where $\{\sigma_{n}\}$ is a matrix basis obeying $\trs{\sigma_{n}} = D\delta_{n0}$ and
$\trs{\sigma_{n}\sigma_{m}^{\dagger}} = D\,\delta_{mn}$\footnote{From these definitions, one identifies the inner product between two elements
$\dket{\rho(t)}$ and $\dket{\lambda(t)}$ as $\dinterpro{\lambda(t)}{\rho(t)}=D\trs{\rho(t)\lambda^{\dagger}(t)}$.}.
Here, we will consider the particular case a superoperator $\Lmath(t)$ whose Jordan decomposition~\cite{Horn:Book}  admits one-dimensional Jordan blocks only, which will be individually
associated with distinct non-crossing time-dependent eigenvalues $\lambda_{\alpha}(t)$.
As provided in Appendix~\ref{ApAdApp}, the adiabatic behavior will follow from the {\it adiabatic quantum condition} (AQC)
\begin{equation}
\max_{t\in[0,\tau]} \xi_{\beta\alpha}(t) \ll 1,
\label{osac}
\end{equation}
where $\xi_{\beta\alpha}(t)$ is the adiabatic parameter, reading
\begin{eqnarray}
\xi_{\beta\alpha}(t) = \left \vert \frac{e^{\int_{0}^{t} \Re\left[\lambda_{\alpha}(t^\prime)-\lambda_{\beta}(t^\prime)\right]dt^\prime} \dinterpro{\Ecal_{\beta}(t)}{\dot{\Dcal}_{\alpha}(t)}}{ \lambda_{\beta}(t)-\lambda_{\alpha}(t) } \right \vert \text{ , } \label{Condition}
\end{eqnarray}
with $\{\dket{\Dcal_{\alpha}(t)}\}$ and  $\{\dbra{\Ecal_{\alpha}(t)}\}$ denoting the right and left eigenstates of $\Lmath(t)$ with eigenvalues $\lambda_{\alpha}(t)$,
respectively, and $\Re [z]$ representing the real part of $z\in \Cmath$.
Eq.~(\ref{osac}) ensures decoupled evolution of Jordan-Lindblad eigenspaces~\cite{Sarandy:05-1}.
The energy scale $|\lambda_{\alpha}(t)-\lambda_{\beta}(t)|$ takes into account both the Hamiltonian spectrum and the decohering rates.
The coherence vector for an arbitrary time $t$ can be expanded as
$\dket{\rho(t)} = \sum_{\alpha} c_{\alpha}(t) e^{\int_{0}^{t} \lambda_{\alpha}(t^\prime)dt^\prime} \dket{\Dcal_{\alpha}(t)}$, where $c_{\alpha}(t)$ are
complex functions of time associated with the right eigenbasis of $\Lmath(t)$.
It is important to highlight that, unlike the closed system case~\cite{Sarandy:04,Amin:09,Tong:07,Jianda-Wu:08,Ambainis:04},
the adiabatic parameter in Eq.~\eqref{Condition} is such that $\xi_{\beta\alpha}(t) \neq \xi_{\alpha\beta}(t)$ due to the presence of the factor
$\eta_{\beta\alpha}(t) \equiv e^{\int_{0}^{t} \Re\left[\lambda_{\alpha}(t^\prime)-\lambda_{\beta}(t^\prime)\right]dt^\prime}$. This means that, if we have $\max_{t\in[0,\tau]} \xi_{\beta\alpha}(t) \ll 1$, then
the dynamics of $\dot{c}_{\beta}(t)$ is decoupled from the coefficient $c_{\alpha}(t)$, but it does not mean that $\dot{c}_{\alpha}(t)$ is decoupled from $c_{\beta}(t)$. Moreover, the real exponential in the factor $\eta_{\beta\alpha}(t)$ is responsible for the general finite optimal time for the adiabatic behavior in open systems~\cite{Sarandy:05-1}. However, we can show that, under decoherence, adiabaticity may still occur in the infinite time limit, as it happens for closed systems, for a class of initial quantum states. Indeed, let $\Lmath(t)$ be a Lindblad superoperator that admits one-dimensional Jordan blocks with distinct eigenvalues $\lambda_{n}(t)$. Assume that the initial state of the system can be written as a superposition of two independent eigenvectors of $\Lmath(0)$ and that there are no eigenvalue crossings in the spectrum of  $\Lmath(t)$. Then, the adiabatic dynamics in open system is achieved for {\it{slowly-varying Lindblad superoperators $\Lmath(t)$, with decoupling of Jordan blocks of $\Lmath(t)$ increasing as $t \rightarrow \infty$.}}  See Appendix~\ref{ApAsympBehav} for a detailed proof of this statement.

\section{Open system AQC for highly oscillating driven fields}

\subsection{Landau-Zener model under dephasing}
In order to investigate the adiabatic behavior of open systems in a controllable experimental realization,
let us consider, as our first example, the single-qubit Landau-Zener Hamiltonian, which is given by
\begin{equation}
H(t) = (\omega_{0}/2)  \sigma_{z} + (\omega_{x}/2) \sin(\omega t)\sigma_{x}  \text{ . }
\label{HamLZ}
\end{equation}
This Hamiltonian presents a resonant configuration when $\omega_{0} \approx \omega$, as shown in Appendix~\ref{ApOscillBehav}. Indeed, it can be used to illustrate
the fact that, due to resonance, several quantitative adiabatic conditions designed for closed systems~\cite{Sarandy:04,Amin:09,Tong:07,Jianda-Wu:08,Ambainis:04,Suter:08} are neither necessary nor sufficient to describe the adiabatic behavior in the model if the analysis is performed in the traditional inertial reference frame~\cite{Hu-18-b}. Concerning decoherence, we consider that the qubit is coupled to a dephasing reservoir evolving under a Lindblad master equation
\begin{eqnarray}
\dot{\rho}(t) = - i [H(t),\rho(t)] + \gamma \left[ \sigma_{z} \rho(t) \sigma_{z} - \rho(t) \right] \text{ , } \label{LindDeph}
\end{eqnarray}
with $\gamma$ denoting a time-independent dephasing rate. By rewriting Eq.~(\ref{LindDeph}) in the superoperator formalism we obtain a $4\times4$-dimensional superoperator $\Lmath (t)$ whose elements are $\Lmath_{nm} (t) = (1/2)\text{Tr} \{- i \sigma_{n} [H(t),\sigma_{m}] + \gamma \sigma_{n}\left[ \sigma_{z} \sigma_{m} \sigma_{z} - \sigma_{m} \right]\}$. So, we have (by adopting the basis $\{\1,\sigma_{x},\sigma_{y},\sigma_{z}\}$)
\begin{eqnarray}
\Lmath (t) = \begin{bmatrix}
0 & 0 & 0 & 0 \\
0 & -2\gamma & - \omega_{0} & \omega_{0}\sin (\omega t) \tan \theta \\
0 & \omega_{0} & -2\gamma & 0 \\
0 & -\omega_{0}\sin (\omega t) \tan \theta & 0 & 0
\end{bmatrix} \text{ . }
\end{eqnarray}

The system is initialized in the ground state of $H(0)$, which is given by $\rho(0) = \ket{1}\bra{1}$. Then, the coherence initial vector $\dket{\rho(0)}$ is written as $\dket{\rho(0)} = \dket{\Dcal_{0}(0)} - \dket{\Dcal_{1}(0)}$, where the eigenvectors of the driving super-operator are $\dket{\Dcal_{0}(t)} = \begin{matrix} [\text{ } 1 & 0 & 0 & 0 \text{ }]^{\text{t}} \end{matrix}$ and $\dket{\Dcal_{1}(t)} = \begin{matrix} [\text{ } 0 & 0 & \sin (\omega t) \tan \theta & 1 \text{ }]^{\text{t}} \end{matrix}$, which are associated with eigenvalues $\lambda_{0} = 0$ and $\lambda_{1} = -2\gamma$, respectively, where $\theta = \arctan [\omega_{x}/\omega_{0}]$ . The other eigenvalues of $\Lmath (t)$ are $\lambda_{2} = -\gamma - \Delta (t)\sec(\theta)/2$ and $\lambda_{2} = -\gamma + \Delta (t)\sec(\theta)/2$, where $\Delta (t)$ is a purely complex number defined as $\Delta^2 (t) = 2 \gamma^2 + \omega_{0}^2 [2 \cos(2 t \omega) \sin^2(\theta) - 3] + (2 \gamma^2 - \omega_{0}^2) \cos(2 \theta)$. Thus, adiabatic dynamics is such that it inhibits transitions from eigenvectors $\dket{\Dcal_{0}(t)}$ and $\dket{\Dcal_{1}(t)}$ to other eigenstates of $\Lmath (t)$. See Appendix~\ref{ApOscillBehav} for a complete characterization of the eigenstates of $\Lmath (t)$.

\subsection{Experimental implementation}

\begin{figure}[t!]
	\centering
	\includegraphics[scale=1.70]{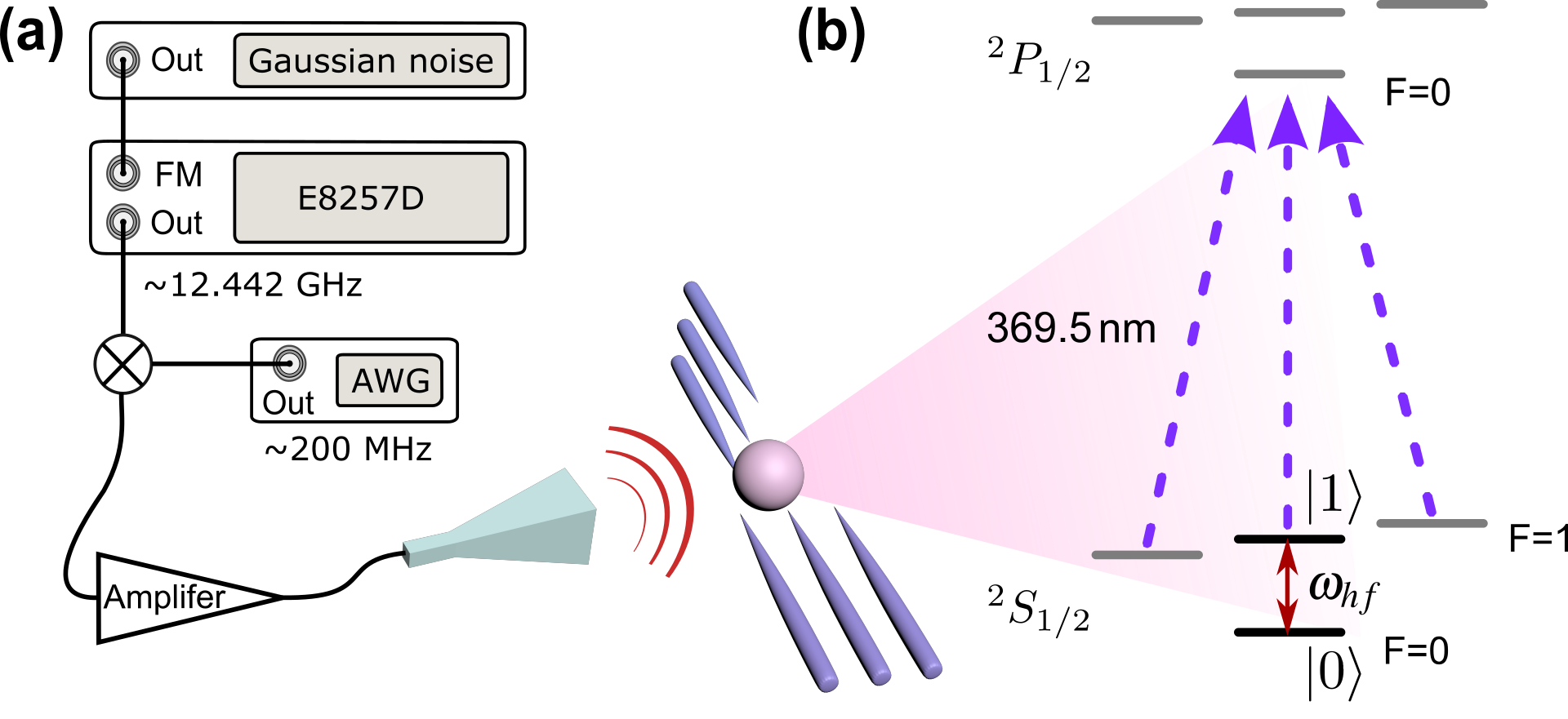}
	\caption{Experimental setup for demonstrating the adiabatic quantum dynamics under decoherence.
		{\color{blue}(a)} Experimental apparatus to generate the operating waveform, where the programmable arbitrary waveform generator (AWG)
		is used to implement the target Hamiltonian, with Gaussian noise working as environment.
		{\color{blue}(b)} The relevant energy spectrum of the $^{171}$Yb$^{+}$ ion and the six needle trap.}
	\label{setup}
\end{figure}

We explore the adiabatic behavior of the resonant Landau-Zener Hamiltonian in a decohering environment both from a theoretical and an experimental point of view. In order to experimentally realize the model, we use a trapped Ytterbium ion ($^{171}$Yb$^{+}$) to implement the dynamics in Eq.~\eqref{LindDeph}, with the experimental setup
shown in Fig.~\ref{setup}. We encode the qubit into two  hyperfine energy levels of the $^{2}S_{1/2}$ ground state, which is denoted by $\ket{0} \equiv\, ^{2}S_{1/2}\, \ket{F=0,m_{F}=0}$ and  $\ket{1}\equiv\, ^{2}S_{1/2}\,\ket{F=1,m_{F}=0}$.
Before the microwave manipulation, we can use a standard optical pumping process to initialize the qubit into the $\ket{0}$ state with $99.9\%$ efficiency.

By using the arbitrary waveform generator (AWG), we coherently drive the hyperfine qubit to implement the target Hamiltonian. This is a well-established
experimental procedure. However, there is no universal way to control the two-level system interacting with an environment. Instead, we have introduced
a new approach, which is based on a Gaussian noise frequency modulation of the $2\pi\times 12.442$ GHz carrier microwave. This can be used to mimic the dephasing
channel with high controllability for an  arbitrary target dephasing rate $\gamma$~\cite{Hu:19}. After the microwave operation, a standard florescence
detection method is used to measure the population of the  $\ket{1}$ state~\cite{Olmschenk:07}.

To verify the effectiveness of the adiabatic condition for open systems, we use the fidelity as a figure of merit. Fidelity can be used as a distinguishability measure between two density matrices $\rho_{1}$ and $\rho_{2}$, being defined by $\Fcal(t) = \Acal[\rho_{1},\rho_{2}]$, where $\Acal[\rho_{1},\rho_{2}] = \text{Tr} [(\rho_{1}^{1/2} \rho_{2} \rho_{1}^{1/2} \text{ })^{1/2} ]$.

\begin{figure}[t!]
	\centering
	\input{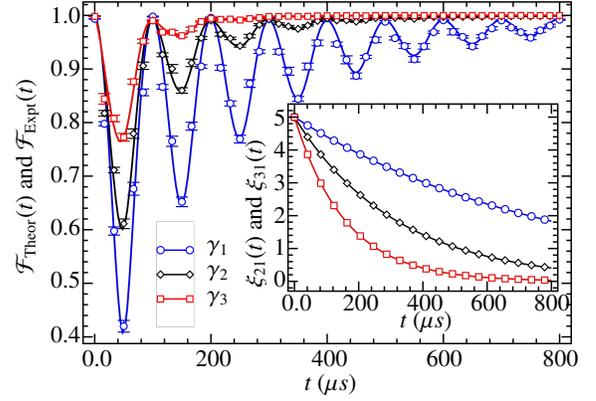}
	\vspace{5cm}
	\caption{Main: Theoretical fidelity $\Fcal_{\text{Theor}}(t)$ (continuum lines) and its experimental counterpart $\Fcal_{\text{Expt}}(t)$ (symbols) as functions of the evolution time $t$
		for the values of the dephasing rate $\gamma_{1}=1256$~Hz (circle), $\gamma_{2}=3141$~Hz (diamond) and $\gamma_{3}=6283$~Hz (square), where the each fidelity is computed concerning state in Eq.~\eqref{rhoad}. Inset: Adiabatic parameters $\xi_{21}(t)$ and $\xi_{31}(t)$ (in multiples of $10^{-4}$). In the inset, the symbols do not represent experimental points. We adopted the values $\omega_{0} = 2\pi$~MHz and $\omega_{x} = 2\pi\times20$~KHz, with resonance obtained through $\omega = \omega_{0}$.}
	\label{Fidel-Osci}
\end{figure}

Concerning the expected adiabatic behavior, we note first that, since the eigenvector $\dket{\Dcal_{0}(t)}=\dket{\Dcal_{0}}$ is time-independent, its dual left eigenstate $\dbra{\Ecal_{0}(t)}=\dbra{\Ecal_{0}}$ is also time-independent, so that we can use the relation $\dinterpro{\Ecal_{\beta}(t)}{\dot{\Dcal}_{\alpha}(t)} = - \dinterpro{\dot{\Ecal}_{\beta}(t)}{\Dcal_{\alpha}(t)}$ to conclude that $\xi_{0\alpha} = \xi_{\alpha0} = 0$. Thus, we do not have transitions from $\dket{\Dcal_{0}}$ to any other state. In addition, as detailed in Appendix~\ref{ApAsympBehav},
$\xi_{\alpha0} = 0$ \textit{is always valid for any open quantum system}. On the other hand, $\xi_{\alpha2}$ is nonvanishing. Its behavior is shown in the inset plot in Fig.~\ref{Fidel-Osci}., which allows us to conclude that the adiabatic approximation shown in Eq.~\eqref{Condition} is indeed satisfied for the range of values for the parameters chosen in the experiment. Since the coefficients $\xi_{\alpha2}$ are small and
decay as a function of time, it is
	then expected that the fidelities increase and approach the value one as time $t\rightarrow \infty$.
Indeed, the adiabatic evolved state can be written as
\begin{eqnarray}
\rho_{\text{ad}}(t) = \frac{1}{2} \left[ \1 - \frac{1}{2}e^{-2\gamma t}\sin(\omega t) \tan(\theta) \sigma_{y} - e^{-2\gamma t}\sigma_{z}\right] \label{rhoad} \text{ . }
\end{eqnarray}

The experimental output data and the theoretical predictions are shown in Fig.~\ref{Fidel-Osci}, where theoretical and experimental process fidelities are, respectively, given by $\Fcal_{\text{Theor}}(t) = \Acal[\rho_{\text{ad}}(t),\rho_{\text{N-sol}}(t)]$ and $\Fcal_{\text{Expt}}(t) = \Acal[\rho_{\text{ad}}(t),\rho_{\text{Expt}}(t)]$, where $\rho_{\text{N-sol}}(t)$ is the numerical solution of Eq.~\eqref{LindDeph} and $\rho_{\text{Expt}}(t)$ is obtained from the standard quantum state tomography. Remarkably, experimental
fidelity is measured as $99.4\%$ in our system~\cite{Hu:18}. In all of the experiments realized in this work, the error bars are obtained from the standard 
deviation associated with 60 000 repetitions of the experiment. For every fidelity, we perform state tomography by measuring the qubit in the three Pauli bases 
($\sigma_{x}$, $\sigma_{y}$, and $\sigma_{z}$)~\cite{James:01}, 
with every basis measured 2000 times and repeated 10 times.

From Fig.~\ref{Fidel-Osci}, notice that fidelities with respect to the adiabatic density operator $\rho_{\text{ad}}(t)$ tend to increase for long evolution times
but they undergo strong oscillations at an intermediate time scale. This is a consequence of the resonance phenomenon. If evolution is not long enough,
the qubit approximately evolves as a closed system, since decoherence effects do not considerably drive the system through a nonunitary evolution. Indeed,
it is known that, in this situation, resonance limits the applicability of the adiabatic theorem (correct predictions for closed systems would require a 
change for a non-inertial frame~\cite{Hu-18-b}). On the other hand, as time increases in an open system scenario,
decoherence plays a relevant role, driving the
system with high fidelities to the adiabatic evolution predicted by the adiabatic approximation established by Eq.~(\ref{osac}). 
This is rather different for closed systems, where resonance challenges
the applicability of the adiabatic theorem for an arbitrary time scale~\cite{Marzlin:04,Suter:08,Hu-18-b}. This result is also in contrast with the general predictions in Refs~\cite{Sarandy:05-1,Sarandy:05-2}, where open system adiabaticity is typically expected to occur at finite time for the case of decomposition in terms of general Jordan blocks.

\section{Adiabatic Deutsch algorithm}

\subsection{Deutsch Hamiltonian under dephasing}

Let us consider now an application in quantum computing by means of the adiabatic Deutsch algorithm in open systems~\cite{Sarandy:05-2}.
The Deutsch problem is a decision problem on whether a dichotomic function $f: x\in \{0,1\} \rightarrow f(x) \in \{0,1\}$ is \textit{constant} [$f(0)=f(1)$] or \textit{balanced}
[$f(0)\ne f(1)$]~\cite{Deutsch:85}. The algorithm is implemented through a time-dependent Hamiltonian governing the adiabatic dynamics of a qubit, which is given by
\begin{eqnarray}
H_{\text{DA}}(t) = - \omega \left[\cos \left(\pi F t/2\tau \right) \sigma_{x} + \sin \left(\pi F t/2\tau \right) \sigma_{y}\right] \text{ , }
\end{eqnarray}
where $F = 1-(-1)^{f(0)+f(1)}$ and $\tau$ is the total evolution time. The qubit is initialized in the state $\ket{\psi(0)} = \ket{+}$. By adiabatically evolving the system, the global behavior of the function $f$ can be
obtained from the final state , which will be $\ket{\psi(\tau)} = \ket{+}$ when $f$ is constant and $\ket{\psi(\tau)} = \ket{-}$ when $f$ balanced.
Here, we will consider the effects of dephasing on the Deutsch algorithm, as provided by the master equation described in Eq.~\eqref{LindDeph}.
\begin{figure}
	\centering
	\input{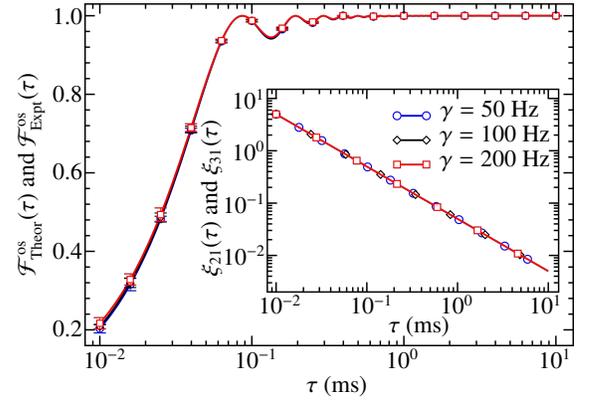}
	\vspace{5cm}
	\caption{Main: Fidelity for the adiabatic open system behavior of the Deutsch algorithm. It is shown $\Fcal^{\text{os}}_{\text{Theor}}(\tau)$ (continuum lines) and its experimental counterpart
	$\Fcal^{\text{os}}_{\text{Expt}}(\tau)$ (symbols) as functions of $\tau$ for several values of the dephasing rate $\gamma$.  Inset: Adiabatic open system parameters $\xi_{21}(\tau)$ and $\xi_{31}(\tau)$.
	In the inset, the symbols do not represent experimental points. They are used just to distinguish the results for each value of $\gamma$. We consider a balanced function with
	$\omega = 2\pi \times 10$~KHz. Due the high controllability and measurement fidelity, some experimental points data are superposed.}
	\label{Fidel-DA}
\end{figure}

The dynamics of the Deutsch algorithm can be obtained from the superoperator
\begin{align}
\Lmath_{\text{DA}}(t) = \begin{bmatrix}
0 & 0 & 0 & 0 \\ 0 & -2 \gamma & 0 & \omega \sin \pi t/\tau \\ 0 & 0 & -2 \gamma & \omega \cos \pi t/\tau \\ 0 & - \omega \sin \pi t/\tau & - \omega \cos \pi t/\tau & 0
\end{bmatrix} ,
\end{align}
where the relevant eigenvalues of $\Lmath_{\text{DA}} (t)$ are given by $\lambda_{0} = 0$ and $\lambda_{1} = -2 \gamma $, with their corresponding eigenvectors reading
$\dket{\Dcal^{\text{DA}}_{0}(s)} = \begin{matrix}
[\text{ } 1 & 0 & 0 & 0\text{ }]^{\text{t}}
\end{matrix}$
and
$\dket{\Dcal^{\text{DA}}_{1}(s)} = \begin{matrix} [\text{ } 0 & \cos ( F \pi s/2) & - \sin ( F \pi s/2) & 0 \text{ }]^{\text{t}} \end{matrix}$, where $s=t/\tau \in [0,1]$ is the normalized evolution time.
The other eigenvalues are $\lambda_{2} = - (\gamma + \Delta) $ and $\lambda_{3} = - (\gamma - \Delta) $, where $\Delta^2 = \gamma^2 - \omega^2$.
Therefore, from this set of eigenvectors, we can identify the initial state $\dket{\psi(0)}$ of the protocol as given by the superposition
$\dket{\rho^{\text{DA}}(0)} = \dket{\Dcal^{\text{DA}}_{0}(0)} + \dket{\Dcal^{\text{DA}}_{1}(0)}$ and the evolved adiabatic matrix density is obtained as (see Appendix~\ref{ApDA})
\begin{equation}
\rho^{\text{os}}_{\text{DA}}(s) = \frac{1}{2} \left[ \1 + e^{-2 \gamma \tau} \cos\left(\frac{F \pi s}{2}\right) \sigma_{x} - e^{-2 \gamma \tau} \sin\left(\frac{F \pi s}{2}\right) \sigma_{y} \right] \text{ , } \label{AdSolDA}
\end{equation}
where the superscript {\it os} stands for {\it open system} in order to indicate that the density operator is obtained from an open system adiabatic dynamics. The above adiabatic solution provides the target state of the algorithm at $s=1$ under the action of the decohering environment as
\begin{eqnarray}
\rho^{\text{os}}_{\text{DA}}(1) = \frac{1}{2} \left[ \1 + e^{-2 \gamma \tau} (-1)^{f(0)+f(1)} \sigma_{x} \right] \text{ . } \label{AdSolDAs1}
\end{eqnarray}
where we use $\cos(F\pi/2) = (-1)^{f(0)+f(1)}$ and $\sin(F\pi/2) = 0$ for any function $f(x) \in \{0,1\}$. The dynamics of the system and its distance with respect to $\rho_{\text{DA}}(s)$ (as measured by the fidelity) are provided in Fig.~\ref{Fidel-DA}, where without loss of generality, we consider the implementation of the algorithm for a balanced function $f(0)=0$ and $f(1)=1$. As before, the theoretical and experimental fidelities of achieving the adiabatic open system dynamics are defined by $\Fcal^{\text{os}}_{\text{Theor}}(\tau) = \Acal[\rho^{\text{os}}_{\text{DA}}(1),\rho_{\text{N-sol}}(\tau)]$ and $\Fcal^{\text{os}}_{\text{Expt}}(\tau) = \Acal[\rho^{\text{os}}_{\text{DA}}(1),\rho_{\text{Expt}}(\tau)]$, respectively, with $\rho_{\text{N-sol}}(\tau)$ given by the numerical solution of Eq.~\eqref{LindDeph} for the Hamiltonian $H_{\text{DA}}(t)$ and $\rho_{\text{Expt}}(t)$ obtained via quantum tomography.

Remarkably, by looking at the adiabatic decohering dynamics of the Deutsch algorithm, it is argued in Ref.~\cite{Sarandy:05-1} that there should be
an optimal time for the adiabatic approximation. While this is true in general, as a consequence of the competition between the adiabatic and the decoherence time scales, here
fidelity increases to its maximum as $\tau \rightarrow \infty$. Indeed, as proved in Appendix~\ref{ApAsympBehav}, this is because
the initial density operator is a superposition of two blocks of $\Lmath_{\text{DA}} (0)$, given that $\Lmath_{\text{DA}} (t)$ admits
one-dimensional Jordan blocks and does not show eigenvalue crossings. As it will be explicitly discussed below, notice also that the high fidelity for the adiabatic behavior as $\tau \rightarrow \infty$ does not imply in a high fidelity for the target state of the algorithm, since the adiabatic solution provides a maximally mixed state in the asymptotic limit.

\begin{figure}[t!]
	\centering
	\input{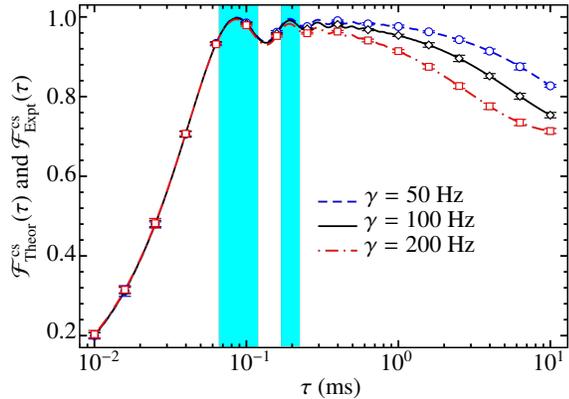}
	\vspace{5cm}
	\caption{Fidelity for the pure target state in the Deutsch algorithm for several values of the dephasing rate $\gamma$.
	Lines represent $\Fcal^{\text{cs}}_{\text{Theor}}(\tau)$ and the dots its experimental counterpart $\Fcal^{\text{cs}}_{\text{Expt}}(\tau)$.
	We consider a balanced function with $\omega = 2\pi \times 10$~KHz. Two high fidelity time windows are highlighted in light green color.}
	\label{Fidel1-DA}
\end{figure}

\subsection{Adiabatic time window for the target state fidelity}

A challenge in the implementation of adiabatic quantum algorithms under decoherence is to obtain a favorable trade-off
between the necessary time to achieve the adiabatic regime and the (restrictive) time scale of decohering effects. 
As previously illustrated, the oscillatory behavior of the fidelity is a common phenomenon in the adiabatic evolution of both 
closed and open quantum systems. In particular, it is possible to identify the first maximum of the fidelity near
to one in different adiabatic protocols and systems~\cite{Santos:18-b,Hu:18,Albash:15,Coulamy:17,Peterson:18}. 
To investigate how useful the fidelity maxima can be, we show in Fig.~\ref{Fidel1-DA} the fidelity of achieving the desired target state 
of the Deutsch algorithm, which is the expected output of the algorithm. Notice that this requires more than simply ensuring open system adiabaticity, since the adiabatic mixed-state density operator 
can only approximate the target pure state desired as the result of the computation process. 
The output for a balanced function $f$ is the pure state $\ket{-}$, which can be represented by the density operator
\begin{equation}
\rho^{\text{out}}_{\text{DA}}(1) = (1/2)(\1 - \sigma_{x}).
\label{douDA}
\end{equation}
Observe that, since we are evolving under decoherence, the target state provided by Eq.~(\ref{douDA}) is distinct of the adiabatic solution given in Eq.~\eqref{AdSolDAs1},
with the adiabatic solution reducing to the target state in the limit of vanishing decoherence.
In order to quantify the success of the algorithm under decoherence, we define the fidelity of the real dynamics with respect to the target output density operator.
The theoretical values are provided by $\Fcal^{\text{cs}}_{\text{Theor}}(\tau) = \Acal[\rho^{\text{out}}_{\text{DA}}(1),\rho_{\text{N-sol}}(\tau)]$, while their
experimental counterparts are $\Fcal^{\text{cs}}_{\text{Expt}}(\tau) = \Acal[\rho^{\text{out}}_{\text{DA}}(1),\rho_{\text{Expt}}(\tau)]$. 
The superscript {\it cs} stands for {\it closed system} in order to indicate the fidelity for the expected target state that 
would be obtained from a closed system adiabatic dynamics. The above adiabatic solution provides the target state of the algorithm at $s=1$ under the action of the decohering environment. 

Notice that the fidelity with respect to the pure target state in Fig.~\ref{Fidel1-DA} decays for long times as the decoherence rate increases. This is in contrast
with the fidelity with respect to the mixed-state adiabatic evolution for open system in Fig.~\ref{Fidel-DA}, which goes to one as $t \rightarrow \infty$.
This is because adiabaticity in open systems is related to
decoupling of Jordan blocks in a nonunitary evolution, which is only approximately equivalent (in the weak coupling regime) to achieving a pure target state
after a computation process. Therefore, by focusing on pure target states, it is possible to see a preferred time-window exhibiting maximum fidelity.
In Fig.~\ref{Fidel1-DA}, we show two windows with high fidelities, which are highlighted in light green color. Notice that these windows also correspond
to high fidelities in Fig.~\ref{Fidel-DA}, which means that adiabaticity in open systems is indeed able to provide the target state of the Deutsch algorithm
with high fidelity for convenient measurement times.

\section{Conclusions}

In this paper we have addressed, from both a theoretical and an experimental point of view,
the resonance sensitiveness and the time optimality of the adiabatic approximation in open systems.
First, we have shown that, distinctly from the closed case, the validity conditions for adiabaticity under decoherence
may be robust against resonant phenomena. Second, we have implemented an adiabatic version of the Deutsch algorithm, analyzing
its run-time optimality in open systems.
In contrast with the general picture previously derived in the literature~\cite{Sarandy:05-1,Sarandy:05-2}, we have shown that the adiabatic
approximation for open systems holds in the asymptotic time limit $t\rightarrow \infty$ in the Deutsch algorithm, which is an arbitrarily
valid consequence of the one-dimensional Jordan
decomposition of the Lindblad superoperator $\Lmath_{\text{DA}}(t)$, the absence of level crossings, and the initialization of the system as
a superposition of only two eigenstates of $\Lmath_{\text{DA}}(t)$.
It is worth mentioning that, if we are interested in maximizing the probability of measuring the target (pure) state as the outcome of the algorithm,
instead of the open system adiabatic mixed-state density
operator, then there is an \textit{optimal time window} for the system measurement (for previously related discussions on this topic, 
see also Refs.~\cite{Steffen:03,Sarandy:05-2,Albash:15,Keck:17}).

Concerning the experimental results, we have reported, to the best of our knowledge, the first experimental investigation of the adiabatic approximation
in a fully controllable open system, where the total evolution time and the decoherence rates can be freely set to verify the quantitative 
validity conditions for the adiabatic behavior.
By using a hyperfine qubit encoded in ground-state energy levels of a trapped Ytterbium ion, all of the theoretical predictions highlighted above have been
successfully realized. In this context, this work generalizes to the open system scenario the experimental analysis of adiabaticity under
resonance implemented for closed systems in Ref.~\cite{Suter:08}. In particular, it has been shown that decoherence may drive the resonant 
quantum system to the open system adiabatic behavior. Moreover, our work also provided a framework to exploit the experimental realization of
adiabaticity in quantum computing under decoherence, tackling features such as optimal run-time of an algorithm, asymptotic time behavior,
competition between adiabatic and relaxation time-scales, outcome fidelities, among others. Scalability of the time window analysis as more qubits are introduced 
and more general features of open systems, such as non-Markovianity, are left as future challenges.

\begin{acknowledgments} This work was supported by the National Key Research and Development Program of China (No. 2017YFA0304100), National Natural Science Foundation of China (Nos. 61327901, 61490711, 11774335, 11734015), Anhui Initiative in Quantum Information Technologies (AHY070000, AHY020100),  Anhui Provincial Natural Science Foundation (No. 1608085QA22), Key Research Program of Frontier Sciences, CAS (No. QYZDY-SSWSLH003), the National Program for Support of Top-notch Young Professionals (Grant No. BB2470000005), the Fundamental Research Funds for the Central Universities (WK2470000026). A.C.S. is supported by Conselho Nacional de Desenvolvimento Cient\'{\i}fico e Tecnol\'ogico (CNPq-Brazil). M.S.S. is supported by CNPq-Brazil (No. 303070/2016-1) and Funda\c{c}\~ao Carlos Chagas Filho de Amparo \`a Pesquisa do Estado do Rio de Janeiro (FAPERJ) (No. 203036/2016). A.C.S. and M.S.S. also acknowledge
financial support in part by the Coordena\c{c}\~ao de Aperfei\c{c}oamento de Pessoal de N\'{\i}vel Superior - Brasil (CAPES) (Finance Code 001) and by the Brazilian
National Institute for Science and Technology of Quantum Information (INCT-IQ).
\end{acknowledgments}

\appendix

\section{Adiabatic approximation: proof of Eqs. \eqref{osac} and \eqref{Condition}} \label{ApAdApp}

The adiabatic approximation in open quantum systems can be appropriately described by the super-operator formalism~\cite{Sarandy:05-1}. To begin with,
let us consider a time-local nonunitary dynamics for a $D$-dimensional quantum system, which reads
\begin{eqnarray}
\dot{\rho}(t) = \Lcal_{t}[\rho(t)] \text{ , }
\label{ApSuperLindEq0}
\end{eqnarray}
where $\Lcal_{t}[\bullet]$ is the dynamical generator. In particular, such equation can be rewritten as
\begin{eqnarray}
\dket{\dot{\rho}(t)} = \Lmath(t)\dket{\rho(t)} \text{ , } \label{ApSuperLindEq}
\end{eqnarray}
where the double ket $\dket{\rho(t)}$ denotes a super-vector (or coherence vector, in case of a single qubit) and $\Lmath(t)$ will be referred as the Lindblad super-operator.
The prefix {\it super} is due to the dimensions of $\dket{\rho(t)}$ and $\Lmath(t)$, namely, $\Lmath(t)$ can be represented by a $D^2 \times D^2 $ matrix and $\dket{\rho(t)}$ by a vector with $D^2$ components, where $D$ is the dimension of the Hilbert space associated with the quantum system. The components $\varrho_{n}(t)$ of the vector $\dket{\rho(t)}$ are obtained from $\varrho_{n}(t) = \trs{\rho(t)\sigma_{n}^{\dagger}}$ and the elements $\Lmath_{mn}(t)$ of $\Lmath(t)$ are written as $\Lmath_{mn}(t) = (1/D)\trs{\sigma_{m}^{\dagger}\Lcal_{t}[\sigma_{n}]}$, where $\{\sigma_{n}\}$ is a matrix basis obeying $\trs{\sigma_{n}\sigma_{m}^{\dagger}} = D\delta_{mn}$. From these definitions, the inner product between two elements $\dket{\rho(t)}$ and $\dket{\lambda(t)}$ is identified as $\dinterpro{\lambda(t)}{\rho(t)}=D\trs{\rho(t)\lambda^{\dagger}(t)}$. In particular, for a single qubit, we get $\Lmath(t)$ as a $4\times 4$ matrix and $\dket{\rho(t)}$ as a  four-vector (vector with 4 components). By writing the density matrix $\rho(t)$ of a single qubit in terms of the components $\varrho_{\alpha}(t)$, we obtain
\begin{eqnarray}
\rho(t) = \frac{1}{2}\left[\1 + \varrho_{x}(t) \sigma_{x}+\varrho_{y}(t)\sigma_{y}+\varrho_{z}(t)\sigma_{z}\right] .
\end{eqnarray}
In general $\Lmath(t)$ does not admit a diagonal form, but it can be decomposed in the Jordan block-diagonal form.
To this end, we define the set of right (quasi-) eigenvectors $\{\dket{\Dcal^{m_\alpha}_{\alpha}(t)}\}$ and its left counterpart $\{\dbra{\Ecal^{m_\alpha}_{\alpha}(t)}\}$,
with $\alpha$ denoting a Jordan block with eigenvalue $\lambda_{\alpha}(t)$ and $m_\alpha$ denoting a (possible) degenerate (quasi-) eigenvector.
The (quasi-) eigenvalue equations are written as~\cite{Sarandy:05-1}
\begin{eqnarray}
\Lmath(t)\dket{\Dcal^{m_\alpha}_{\alpha}(t)} &=& \lambda_{\alpha}(t)\dket{\Dcal^{m_\alpha}_{\alpha}(t)} + \dket{\Dcal_{\alpha}^{m_{\alpha}-1}(t)} \text{ , } \\
\dbra{\Ecal^{m_\alpha}_{\alpha}(t)}\Lmath(t) &=& \lambda_{\alpha}(t)\dbra{\Ecal^{m_\alpha}_{\alpha}(t)} + \dbra{\Ecal_{\alpha}^{m_{\alpha}+1}(t)} \text{ . }
\end{eqnarray}
If the superoperator $\Lmath(t)$ admits an one-dimensional Jordan block decomposition, we can simplify the above equations as
\begin{eqnarray}
\Lmath(t)\dket{\Dcal_{\alpha}(t)} &=& \lambda_{\alpha}(t)\dket{\Dcal_{\alpha}(t)} \text{ , } \label{L1D1} \\
\dbra{\Ecal_{\alpha}(t)}\Lmath(t) &=& \lambda_{\alpha}(t)\dbra{\Ecal_{\alpha}(t)} \text{ . }
\label{L1D2}
\end{eqnarray}
We will assume superoperators $\Lmath(t)$ that admit the Jordan decomposition given by Eqs.~(\ref{L1D1}) and~(\ref{L1D2}).
Thus, the solution of Eq.~\eqref{ApSuperLindEq} can be represented as
\begin{eqnarray}
\dket{\rho(t)} = \sum_{\alpha} c_{\alpha}(t) e^{\int_{0}^{t} \lambda_{\alpha}(t^\prime)dt^\prime} \dket{\Dcal_{\alpha}(t)} \text{ , } \label{ApSolAnsatz}
\end{eqnarray}
where $c_{\alpha}(t)$ are coefficients to be determined. Now, using the above \textit{ansatz} in Eq.~\eqref{ApSuperLindEq}, we find
\begin{eqnarray}
\sum_{\alpha} c_{\alpha}(t) e^{\int_{0}^{t} \lambda_{\alpha}(t^\prime)dt^\prime} \dket{\Dcal_{\alpha}^{m_{\alpha}-1}(t)} &=& \sum_{\alpha} \dot{c}_{\alpha}(t) e^{\int_{0}^{t} \lambda_{\alpha}(t^\prime)dt^\prime} \dket{\Dcal_{\alpha}(t)} \nonumber \\
&+& \sum_{\alpha} c_{\alpha}(t) e^{\int_{0}^{t} \lambda_{\alpha}(t^\prime)dt^\prime} \dket{\dot{\Dcal}_{\alpha}(t)} \text{ , } \nonumber
\end{eqnarray}
so that each $c_{\beta}(t)$ evolves as
\begin{eqnarray}
\dot{c}_{\beta}(t) e^{\int_{0}^{t} \lambda_{\beta}(t^\prime)dt^\prime} &=& c_{\beta}(t) \dinterpro{\Ecal_{\beta}(t)}{\dot{\Dcal}_{\beta}(t)}e^{\int_{0}^{t} \lambda_{\beta}(t^\prime)dt^\prime} \nonumber \\
&+& \sum_{\alpha \neq\beta} c_{\alpha}(t) \dinterpro{\Ecal_{\beta}(t)}{\dot{\Dcal}_{\alpha}(t)}e^{\int_{0}^{t} \lambda_{\alpha}(t^\prime)dt^\prime} \label{S7} \text{ . }
\end{eqnarray}
We can rewrite Eq.~(\ref{S7}) as
\begin{equation}
\frac{d}{dt} \left[c_{\beta}(t) e^{ - \int_{0}^{t} \varkappa_{\alpha\alpha}(t^\prime)dt^\prime} \right] = \sum_{\alpha \neq\beta} c_{\alpha}(t) \varkappa_{\beta\alpha}(t) e^{\int_{0}^{t} (\lambda_{\alpha}(t^\prime)- \lambda_{\beta}(t^\prime)) dt^\prime} \label{S8} \text{ , }
\end{equation}
where we defined $\varkappa_{\beta\alpha}(t) =  \dinterpro{\Ecal_{\beta}(t)}{\dot{\Dcal}_{\alpha}(t)}$. Thus, to achieve the adiabatic behavior,
the right-hand-side of Eq.~(\ref{S8}) must vanish, so that we get
\begin{eqnarray}
c_{\beta}(t) \approx c_{\beta}(0)e^{\int_{0}^{t} \varkappa_{\alpha\alpha}(t^\prime)dt^\prime} \text{ . }
\end{eqnarray}
Thus, a sufficient condition to obtain such result arises by imposing
\begin{equation}
\max _{t\in[0,\tau]} \left \vert \frac{e^{\int_{0}^{t} (\lambda_{\alpha}(t^\prime)- \lambda_{\beta}(t^\prime))dt^\prime} \varkappa_{\beta\alpha}(t)}{ \lambda_{\beta}(t)-\lambda_{\alpha}(t) } \right \vert  \ll 1, \label{Condition1Ap}
\end{equation}
where the eigenvalue gap $|\lambda_{\alpha}(t)-\lambda_{\beta}(t)|$ appears as a natural energy scale to set the adiabatic behavior of the system.
In general, since $\Lmath(t)$ is not Hermitian, each eigenvalue $\lambda_{\alpha}(t)$ can be decomposed as
$\lambda_{\alpha}(t) = \Re \lambda_{\alpha}(t) + i \Im \lambda_{\alpha}(t)$. Then, Eq.~(\ref{Condition1Ap}) becomes
\begin{eqnarray}
\max _{t\in[0,\tau]} \left \vert \frac{e^{\int_{0}^{t} \Re\left[\lambda_{\alpha}(t^\prime)-\lambda_{\beta}(t^\prime)\right]dt^\prime} \varkappa_{\beta\alpha}(t)}{ \lambda_{\beta}(t)-\lambda_{\alpha}(t) } \right \vert  \ll 1 \text{ , } \label{ConditionAp}
\end{eqnarray}
Hence, Eq.~(\ref{ConditionAp}) leads to the adiabatic coefficient in Eq.~\eqref{Condition}, which reads
\begin{eqnarray}
\xi_{\beta\alpha}(t) = \left \vert \frac{e^{\int_{0}^{t} \Re\left[\lambda_{\alpha}(t^\prime)-\lambda_{\beta}(t^\prime)\right]dt^\prime} \dinterpro{\Ecal_{\beta}(t)}{\dot{\Dcal}_{\alpha}(t)}}{ \lambda_{\beta}(t)-\lambda_{\alpha}(t) } \right \vert \text{ , } \label{ApCondition}
\end{eqnarray}

\section{Asymptotic adiabatic dynamics for open systems} \label{ApAsympBehav}

By using the master equation~(\ref{ApSuperLindEq0}), it follows that the superoperator $\Lcal_{t}[\bullet]$ satisfies $\trs{\Lcal_{t}[\rho(t)]} = 0$, since $\trs{\dot{\rho}(t)} = 0$.
Thus, if we consider the basis $\{\sigma_{n}\}$, with $\sigma_{0}=\1$, the first row of the matrix representation of $\Lmath(t)$ is vanishing.
In fact, by adopting that the matrix elements of $\Lmath(t)$ are written as $\Lmath_{mn}(t) = (1/D)\trs{\sigma_{m}^{\dagger}\Lcal_{t}[\sigma^{n}]}$, we have
\begin{eqnarray}
\Lmath_{0n}(t) = \frac{1}{D}\trs{\Lcal_{t}[\sigma^{n}]} = 0 \text{ . }
\end{eqnarray}
Then $\Lmath(t)$ has at least one eigenvalue zero with eigenvector constant. We further assume that $\Lmath(t)$ admits one-dimensional Jordan decomposition and
that there are no eigenvalue crossings in the spectrum of $\Lmath(t)$.
If we suitably order the basis so that the first eigenvalue is $\lambda_{0} = 0$, we get the associated eigenvector as
\begin{eqnarray}
\dket{\Dcal_{0}} = \begin{bmatrix}
1 & 0 & 0 & \cdots & 0 & 0
\end{bmatrix}^t \text{ . }
\end{eqnarray}
As an immediate result, it follows that $\dbra{\Ecal_{0}} = \dket{\Dcal_{0}}^{t}$. In addition, such vector is associated with the maximally mixed state $(1/D)\1$.
In fact, the elements $\varrho_{n}(t)$ of $\dket{\rho(t)}$ are given by $\varrho_{n}(t) = \trs{\sigma_{n}^{\dagger}\rho(t)}$, so that in the basis $\{\1,\sigma_{n}\}$ we have
\begin{eqnarray}
\varrho_{n}(t) &=& \frac{1}{D}\trs{\sigma_{k}^{\dagger} \1} = \delta_{0n} \text{ . }
\end{eqnarray}
where we use that $\trs{\sigma_{k}^{\dagger}\sigma_{n}} = D \delta_{kn}$ and $\trs{\sigma_{n}} = D\delta_{n0}$. Therefore, by writing the density matrix as
\begin{eqnarray}
\rho(t) = \frac{1}{D}\left[\1 + \sum_{n=1}^{D-1} \varrho_{n}(t) \sigma_{n} \right] \text{ , }
\end{eqnarray}
it is possible to conclude that any physical initial state $\dket{\rho(0)}$ \textit{must} be written as a combination of $\dket{\Dcal_{0}}$ and other vectors $\dket{\Dcal_{\beta\neq 0}(t)}$. Hence, the initial state can be generally written as
\begin{eqnarray}
\dket{\rho(0)} = c_0(0) \dket{\Dcal_{0}} +\sum\nolimits_{\beta \neq 0} c_{\beta}(0) \dket{\Dcal_{\beta}(0)} \text{ , } \label{ApInitState}
\end{eqnarray}
where $c_0(0)=1$ and $c_{\beta}(0)$ are general complex coefficients.

The dynamics of the vanishing eigenvalue subspace can be studied from Eq.~\eqref{S7}. Indeed, let us write
\begin{eqnarray}
\dot{c}_{0}(t) &=& \sum_{\beta \neq0} c_{\beta}(t) \dinterpro{\Ecal_{0}}{\dot{\Dcal}_{\beta}(t)}e^{\int_{0}^{t} \lambda_{\beta}(t^\prime)dt^\prime} \text{ . }\label{Apc0dot}
\end{eqnarray}
where we already used $\lambda_{0}(t) = 0$ and $\dinterpro{\Ecal_{\beta}(t)}{\dot{\Dcal}_{0}(t)} = 0$. Now, using that the supervectors $\dket{\dot{\Dcal}_{\beta}(t)}$ and $\dbra{\Ecal_{\beta}(t)}$ satisfy $\dinterpro{\Ecal_{\alpha}(t)}{{\Dcal}_{\beta}(t)} = \delta_{\beta \alpha}$ for any $\beta$ and $\alpha$, we have
\begin{eqnarray}
\frac{d}{dt}\dinterpro{\Ecal_{\alpha}(t)}{\Dcal_{\beta}(t)} = \dinterpro{\dot{\Ecal}_{\alpha}(t)}{\Dcal_{\beta}(t)}+ \dinterpro{\Ecal_{\alpha}(t)}{\dot{\Dcal}_{\beta}(t)} = 0 \text{ , } \nonumber
\end{eqnarray}
and consequently we conclude that
\begin{eqnarray}
\dinterpro{\dot{\Ecal}_{\alpha}(t)}{\Dcal_{\beta}(t)} = - \dinterpro{\Ecal_{\alpha}(t)}{\dot{\Dcal}_{\beta}(t)} \text{ . }
\end{eqnarray}
By using this result in Eq.~\eqref{Apc0dot}, we get $\dot{c}_{0}(t) = 0$, which then implies in $c_0(t)=1, \, \forall t$.

Concerning the dynamics of the remaining eigenstates of $\Lmath(t)$, let us assume that
the initial state is such that a single Jordan block is populated in addition to the block associated with $\lambda_0(t)$,
i.e., $c_\eta(0) \ne 0$ for a single $\eta \in \left[1,D^2-1\right]$.
We then start by looking at the adiabatic condition in Eq.~\eqref{ApCondition}.
The parameter $\xi_{\beta\alpha}(t)$ tells us whether or not
$\beta$ can evolve decoupled from $\alpha$, but it does not provide any information whether $\alpha$ can evolve decoupled from $\beta$.
Assume that $\max_{t\in[0,\tau]}\xi_{\beta\alpha}(t) \ll 1$ asymptotically in time ($t \rightarrow \infty$),
due to the fact that $\Re\left[\lambda_{\alpha}(t)-\lambda_{\beta}(t)\right] < 0, \, \forall \alpha,t$.
Provided the absence of level crossings as a function of time,
this condition selects the largest eigenvalue $\lambda_{\beta}(t)$.
Then, we write $\dot{c}_{\beta}(t) =  b(t) c_{\beta}(t)$, with $b(t)$ denoting a complex coefficient. The solution reads $c_{\beta}(t) = c_{\beta}(0)e^{\int_{0}^{t}b(t^\prime)dt^\prime}$.
On the other hand, once the parameter $\xi_{\alpha\beta}(t)$ does not necessarily satisfy $\max_{t\in[0,\tau]}\xi_{\alpha\beta}(t) \ll 1$, we generically write
$\dot{c}_{\alpha}(t) = \sum_{\alpha^\prime \neq \beta} a_{\alpha^\prime} (t) c_{\alpha^\prime}(t) + a_\beta(t) c_{\beta}(t)$, with $a_{\alpha^\prime}(t)$ and $a_{\beta}(t)$
complex coefficients. However, by imposing $c_{\beta}(0) = 0$,
it yields ${c}_{\beta}(t) = 0$ and consequently
$\dot{c}_{\alpha}(t) = \sum_{\alpha^\prime \neq \beta} a_{\alpha^\prime} (t) c_{\alpha^\prime}(t)$. By iteratively applying this argument after decoupling $\lambda_\beta(t)$,
we finally obtain that $\dot{c}_{\eta}(t) = a_{\eta}(t) c_{\eta}(t)$, stopping as $\lambda_\eta(t)$ becomes the largest eigenvalue of the remaining set.
Thus, by assuming a single $c_\eta(0) \ne 0$ in the initial state, which is equivalent to assuming an initial superposition of only two Jordan blocks, an adiabatic dynamics
for $t\rightarrow \infty$ is always achieved, reading
\begin{eqnarray}
\dket{\rho(t)} = \dket{\Dcal_{0}} + c_{\eta}(t) \dket{\Dcal_{\eta}(t)} \text{ . }
\end{eqnarray}

\section{Open system AQC for highly oscillating driven fields} \label{ApOscillBehav}

Let us consider a Landau-Zener type Hamiltonian given by
\begin{eqnarray}
H(t) = \frac{\omega_{0}}{2}  \sigma_{z} + \frac{\omega_{\text{x}}}{2} \sin(\omega t)\sigma_{x}  \text{ .}
\label{HLZAPPE}
\end{eqnarray}
We can rewrite Eq.~(\ref{HLZAPPE}) as
\begin{eqnarray}
H(t) = \frac{\omega_{0}}{2} \left[ \sigma_{z} + \tan \theta \sin(\omega t)\sigma_{x} \right] \text{ , } \label{ApResH}
\end{eqnarray}
where $\theta = \arctan [\omega_{\text{x}}/\omega_{0}]$. It is not obvious that the Hamiltonian $H(t)$ in Eq.~\eqref{ApResH} exhibits a resonant behavior. In order to see this fact, let us define a time-dependent oscillating frame $R(t) = e^{-i\frac{\omega}{2}t\sigma_{z}}$. In this oscillating frame, the Hamiltonian is given by
\begin{equation}
H_{\text{R}}(t) = R(t) H(t) R^{\dagger}(t) + i R(t)\dot{R}^{\dagger}(t) = \frac{\omega_{0} - \omega}{2} \sigma_{z} + \frac{f(t)}{2} \sigma_{y}
\text{ , }  \label{ApResHR}
\end{equation}
where $ f(t) = e^{i t \omega} \omega_{0} \sin (t \omega) \tan \theta$. Notice then that we can get a resonant situation if we choose $\omega_{0} \approx \omega$.
Concerning the Lindblad superoperator in the superoperator formalism, Eq.~(\ref{LindDeph}) of the main text implies that
\begin{eqnarray}
\Lmath (t) = \begin{bmatrix}
0 & 0 & 0 & 0 \\
0 & -2\gamma & - \omega_{0} & \omega_{0}\sin (\omega t) \tan \theta \\
0 & \omega_{0} & -2\gamma & 0 \\
0 & -\omega_{0}\sin (\omega t) \tan \theta & 0 & 0
\end{bmatrix} \text{ , }
\end{eqnarray}
whose eigenvectors are given by
\begin{subequations}
	\begin{eqnarray}
	\dket{\Dcal_{0}(t)} &=& \begin{matrix} [\text{ } 1 & 0 & 0 & 0 \text{ }]^{\text{t}} \end{matrix} \text{ , } \\
	\dket{\Dcal_{1}(t)} &=& \begin{matrix} [\text{ } 0 & 0 & \sin (\omega t) \tan \theta & 1 \text{ }]^{\text{t}} \end{matrix} \text{ , } \\
	\dket{\Dcal_{2}(t)} &=& \begin{matrix} [\text{ } 0 & -\lambda_{4}(t)\omega_{0} & 1 & -\sin \omega t \tan \theta \text{ }]^{\text{t}} \end{matrix} \text{, \, \, \, } \\
	\dket{\Dcal_{3}(t)} &=& \begin{matrix} [\text{ } 0 & -\lambda_{3}(t)\omega_{0} & 1 & -\sin \omega t \tan \theta \text{ }]^{\text{t}} \end{matrix} \text{. \, \, \, }
	\end{eqnarray}
	\label{ApEigenVect}
\end{subequations}
The eigenvectors above are associated with the eigenvalues $\lambda_{0} = 0$, $\lambda_{1} = -2\gamma$, $\lambda_{2}(t) = -\gamma - \Delta (t)\sec(\theta)/2$ and
$\lambda_{3}(t) = -\gamma + \Delta (t)\sec(\theta)/2$, respectively, where we have defined $\Delta^2 (t) = 2 \gamma^2 + \omega_{0}^2 [2 \cos(2 t \omega) \sin^2(\theta) - 3] + (2 \gamma^2 - \omega_{0}^2) \cos(2 \theta)$.
The system is initialized in the ground state of $H(0)$ as $\ket{1}$ so that, in terms of the Pauli matrices, the initial state is given by
\begin{eqnarray}
\rho(0) = \ket{1}\bra{1} = \frac{1}{2} \left[ \1 - \sigma_{z}\right] \text{ . }
\end{eqnarray}
The initial state in the superoperator formalism is then $\dket{\rho(0)} = \begin{matrix} [\text{ } 1 & 0 & 0 & -1 \text{ }]^{\text{t}} \end{matrix}$. Therefore,
from Eqs.~\eqref{ApEigenVect}, we get
\begin{eqnarray}
\dket{\rho(0)} = \dket{\Dcal_{0}(0)} - \dket{\Dcal_{1}(0)} \text{ . }
\end{eqnarray}
By imposing now an adiabatic evolution under $\Lmath(t)$ we obtain that the evolved state is
\begin{eqnarray}
\dket{\rho(t)} &=& \dket{\Dcal_{0}(t)} - e^{\int_{0}^{t}\lambda_{1}(t^\prime)-\lambda_{0}(t^\prime)dt^\prime} \dket{\Dcal_{1}(t)} \nonumber \\
&=& \dket{\Dcal_{0}(t)} - e^{-2\gamma t} \dket{\Dcal_{1}(t)} \nonumber \\
&=& \begin{matrix} [\text{ } 1 & 0 & -e^{-2\gamma t}\sin(\omega t) \tan(\theta)/2 & -e^{-2\gamma t} \text{ }]^{\text{t}} \end{matrix} \text{ . }
\label{frAPPE}
\end{eqnarray}
From Eq.~(\ref{frAPPE}), the adiabatic density matrix can then be written as
\begin{eqnarray}
\rho(t) = \frac{1}{2} \left[ \1 - \frac{1}{2}e^{-2\gamma t}\sin(\omega t) (\theta) \sigma_{y} - e^{-2\gamma t}\sigma_{z}\right] \text{ . }
\end{eqnarray}

\section{Adiabatic Deutsch algorithm under dephasing channel} \label{ApDA}

As it has been discussed in Ref. \cite{Sarandy:05-2}, an adiabatic Hamiltonian capable of implementing the Deutsch algorithm can be written as
\begin{eqnarray}
H(t) = - \omega \cos \left(\frac{\pi F t}{2\tau} \right) \sigma_{x} - \omega \sin \left(\frac{\pi F t}{2\tau} \right) \sigma_{y} \text{ , }
\label{HDAAPP}
\end{eqnarray}
where where $\tau$ is the total evolution time and $F = (-1)^{f(0)} - (-1)^{f(1)}$.
Thus, if we drive the system by Eq.~(\ref{HDAAPP}), the system is initialized in the state $\ket{\psi(0)} = \ket{+}$ and it adiabatically evolves to $\ket{\psi(\tau)} = \ket{+}$ ($\ket{\psi(\tau)} = \ket{-}$) when $f$ is constant (balanced). Again, we will study the dynamics given by the master equation \eqref{LindDeph} of the main text. In order to analyze the adiabatic behavior of the system, we then consider the Lindblad equation in the superoperator formalism \cite{Sarandy:05-1}. By writing the density operator $\rho(t)$ in terms of its
components ${\varrho}_{n}(t)$, we obtain
\begin{eqnarray}
\rho(t) = \frac{\1 + {\varrho}_{x}(t) \sigma_{x}+{\varrho}_{y}(t)\sigma_{y}+{\varrho}_{z}(t)\sigma_{z}}{2} \text{ . }
\end{eqnarray}
Then, we can write $\dket{\rho(t)} = [\begin{matrix}
1 & \varrho_{x}(t) & \varrho_{y}(t) & \varrho_{z}(t)
\end{matrix}]^{T}$ as the cohering supervector, whose components are $\varrho_{n}(t) = \trs{\rho(t) \sigma_{n}}$. The Lindblad superoperator $\Lmath (t)$ has elements
computed from $\Lmath_{nm} (t) = \frac{1}{2}\text{Tr} \{- i \sigma_{n} [H(t),\sigma_{m}] + \gamma \sigma_{n}\left[ \sigma_{z} \sigma_{m} \sigma_{z} - \sigma_{m} \right]\}$. Thus
\begin{eqnarray}
\Lmath (t) =
\begin{bmatrix}
0 & 0 & 0 & 0 \\ 0 & -2 \gamma & 0 & q(t) \\ 0 & 0 & -2 \gamma & - r(t) \\ 0 & - q(t) & r(t) & 0
\end{bmatrix} \text{ , }
\end{eqnarray}
where $r(t) = -\cos(\pi F s/2)$ and $q(t) = \sin(\pi F s/2)$. The set of eigenvalues of $\Lmath (t)$ is composed of $\lambda_{0} = 0$, $\lambda_{1} = -2 \gamma $, $\lambda_{2} = - (\gamma - \Delta) $ and $\lambda_{3} = - (\gamma + \Delta) $, where $\Delta = \sqrt{\gamma^2 - \omega^2}$. Therefore, $\Lmath (t)$ is composed by one-dimensional Jordan blocks. In addition, the eigenvectors of $\Lmath (t)$ are
\begin{eqnarray}
\dket{\Dcal_{0}(s)} &=& \begin{matrix}
[\text{ } 1 & 0 & 0 & 0\text{ }]^{\text{t}}
\end{matrix} \text{ , } \\
\dket{\Dcal_{1}(s)} &=& \begin{matrix} [\text{ } 0 & \cos\left(\frac{F \pi s}{2}\right) & - \sin\left(\frac{F \pi s}{2}\right) & 0 \text{ }]^{\text{t}} \end{matrix} \text{ , } \label{D1} \\
\dket{\Dcal_{2}(s)} &=& \begin{matrix}
[ \text{ } 0 & \eta^{-1}_{3} \sin\left(\frac{F \pi s}{2}\right) & \eta^{-1}_{3} \cos\left(\frac{F \pi s}{2}\right) & 1 \text{ } ]^{\text{t}}
\end{matrix} \text{ , } \\
\dket{\Dcal_{3}(s)} &=& \begin{matrix}
[ \text{ } 0 & \eta^{-1}_{3} \sin\left(\frac{F \pi s}{2}\right) & \eta^{-1}_{3} \cos\left(\frac{F \pi s}{2}\right) & 1 \text{ } ]^{\text{t}}
\end{matrix} \text{ , }
\end{eqnarray}
where we have defined $\eta_{3} = - \lambda_{3}/\omega$ and $s = t/\tau$. Therefore, in this formalism, the initial state $\rho(0) = (\1 + \sigma_{x})/2$ is represented by the cohering supervector $\dket{\rho(0)} = [\begin{matrix}
1 & 1 & 0 & 0
\end{matrix}]^{T}$. It is important to highlight that, from the set of eigenvectors of $\Lmath (t)$, we can see that the initial state can be written as
\begin{eqnarray}
\dket{\rho(0)} = c_{0}(0) \dket{\Dcal_{0}(0)} + c_{1}(0) \dket{\Dcal_{1}(0)} \text{ , }
\end{eqnarray}
where $c_{0}(0) = c_{1}(0) = 1$. Therefore, the adiabatic approximation for open systems \cite{Sarandy:05-1} states that our system will undergo an adiabatic dynamics given by
\begin{eqnarray}
\dket{\rho(s)} = c_{0}(0)e^{\vartheta_{0}(s)} \dket{\Dcal_{0}(s)} + c_{1}(0)e^{\vartheta_{1}(s)} \dket{\Dcal_{1}(s)} \text{ , }
\end{eqnarray}
where $\vartheta_{0}(s) = 0$ and $\vartheta_{1}(s) = -2 \gamma \tau s$. In matrix form, the evolved density matrix reads as 
\begin{equation}
\rho(s) = \frac{1}{2} \left[ \1 + e^{-2 \gamma \tau} \cos\left(\frac{F \pi s}{2}\right) \sigma_{x} - e^{-2 \gamma \tau} \sin\left(\frac{F \pi s}{2}\right) \sigma_{y} \right] \text{ . }
\end{equation}

Therefore, from the above equation and by using that $\cos(F\pi/2) = (-1)^{f(0)+f(1)}$ and $\sin(F\pi/2) = 0$ for any function $f(x) \in \{0,1\}$,
the density matrix associated with the output state reads
\begin{eqnarray}
\rho(1) = \frac{1}{2} \left[ \1 + e^{-2 \gamma \tau} (-1)^{f(0)+f(1)} \sigma_{x} \right] \text{ . }
\end{eqnarray}


\end{document}